\documentclass{article}
\usepackage{spconf,amsmath,graphicx}
\usepackage{amsfonts}
\usepackage{booktabs}
\usepackage{siunitx}
\usepackage{tabu}
\usepackage{lipsum}
\usepackage{CJKutf8}
\usepackage{url}
\usepackage{hyperref}

\title{Improving Noisy Student Training on Non-target Domain Data for Automatic Speech Recognition }
\name{Yu Chen$^{1,2}$, Wen Ding$^{2\dagger}$ and Junjie Lai$^2$\thanks{Wen Ding$^\dagger$ is the corresponding author. This work is done during Yu Chen's internship at NVIDIA. Thanks to Yuekai Zhang and Hainan Xu for helpful suggestions. This work has been open-sourced into \href{https://github.com/wenet-e2e/wenet/tree/main/examples/aishell/NST}{\textbf{WeNet toolkit}}. }}
\address{$^1$The University of Hong Kong, Hong Kong China \\
  $^2$NVIDIA, Shanghai China  \\
\url{u3603296@connect.hku.hk} , \url{{ wend, julienl}@nvidia.com} }
\begin{document}

%
\maketitle
\begin{abstract}
Noisy Student Training (NST) has recently demonstrated extremely strong performance in Automatic Speech Recognition (ASR). 
In this paper, we propose a data selection strategy named \textit{LM Filter} to improve the performance of NST on non-target domain data in ASR tasks. 
Hypotheses with and without a Language Model are generated and the CER differences between them are utilized as a filter threshold.
Results reveal that significant improvements of 10.4\% compared with no data filtering baselines. We can achieve 3.31\% CER in AISHELL-1 test set, which is best result from our knowledge without any other supervised data. 
We also perform evaluations on the supervised 1000 hour AISHELL-2 dataset and competitive results of 4.73\% CER can be achieved.
\end{abstract}
\begin{keywords}
Data Selection Strategy, Noisy Student Training, Speech Recognition, Semi-supervised Learning
\end{keywords}
\section{Introduction}
\label{sec:intro}




In recent years, Semi-Supervised Learning (SSL) has attracted a lot of research interest 
in many fields of deep learning, such as Automatic Speech Recognition (ASR) \cite{ssl_asr0,ssl_asr1,synnaeve2020endtoend} , Computer Vision \cite{ssl_cv0,ssl_cv1,xie2020self} and Natural Language Processing \cite{ssl_nlp0,ssl_nlp1,ssl_nlp2}. 
Among these methods, Noisy Student Training (NST) has recently demonstrated extremely strong performances in Image Classification \cite{xie2020self} by introducing noise and randomness into traditional Teacher-student Learning \cite{teacher_student0,teacher_student1}. This method further demonstrates its robustness in the ASR field \cite{park2020improved, doutre2021improving, nst_asr4}. After combing with pre-train methods \cite{baevski2020wav2vec}, NST is shown to be a vital component for achieving SOTA results on a number of datasets, e.g. Librispeech \cite{NST_ASR2_pushing_limits}.

However, NST has not been widely investigated in ASR tasks when the domain of the supervised data does not match the unsupervised data. Noise and domain play an important role in ASR \cite{noise_domain0} and the abundant unsupervised data from social media may not always match the domain of the desired task. Thus, proper data selection techniques are required to remove noise and select data that is close to the target domain \cite{lu22_interspeech}. 
The most common filter in ASR is the Confidence Score that selects the most trustworthy transcriptions based on confidence estimation and threshold \cite{confidence_score0, confidence_score1, confidence_score2}. However, this method is not always promising in scenarios with large amount of unlabelled data with domain mismatches.
Another recent unsupervised data selection technique is investigated in \cite{lu22_interspeech}, where a contrastive Language Model is applied as a data selector to better improve the target-domain ASR task.

In this paper, we propose a novel data selection strategy named \textit{LM Filter} which can utilize model differences to filter more valuable non-target domain data to improve the performance of NST. 
We leverage concept of contrastive LM and data selection method in \cite{zheng2022scaling}. Our \textit{LM Filter} is based on hypotheses from LM to gradually remove noisy data inside each iteration of NST method. The filter condition is relaxed through the NST iteration to make the model advance gradually in due order. 
This method has the following benefits:
\begin{itemize}
    \item No additional data selection models are required. Model differences can be obtained from different decoding strategies (e.g. with/without LM).
    \item Label is not required to perform the data selection and it is totally unsupervised.
    \item Less time and resources are utilized to run the NST method and it can converge faster in fewer iterations.
\end{itemize}

Experiments on AISHELL-1 \cite{aishell1} as supervised data and WenetSpeech \cite{zhang2022WenetSpeech} as unsupervised data indicate a significant improvement of 10.4\% comparing with no data filtering baselines. 
When combined AISHELL-2 \cite{aishell2} and WenetSpeech as unsupervised data, 3.31\% character error rate(CER) is achieved on AISHELL-1 test set, which is the best result from our knowledge without any other supervised data on this test set. 
\textit{LM Filter} further demonstrates its robustness in larger dataset such as AISHELL-2 (supervised) and WenetSpeech (unsupervised) to achieve promising result of 4.73\% which has 13.6\% improvement comparing with the baseline. 


The rest of the paper is organized as follows. Section 2 briefly introduces the basic concepts and methods of NST in ASR. Our proposed data selection strategy LM Filter will be included in section 3. Experiment details are introduced in Section 4. Eventually we give our conclusions in section 5.





\section{Noisy Student Training for ASR}  
\label{sec:NST}

\begin{figure}[htb]
\centering
\includegraphics[width=8.5cm]{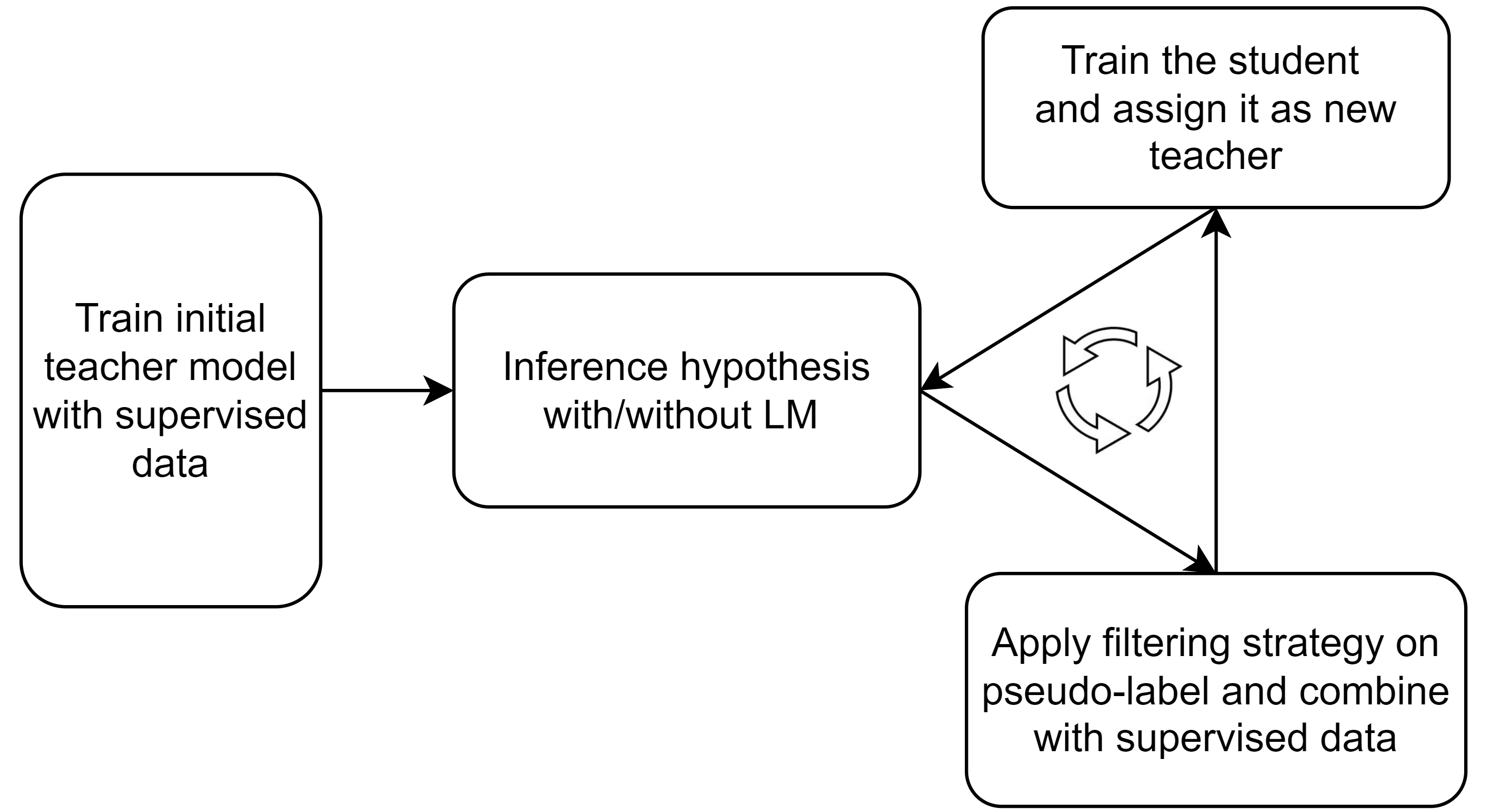}
\caption{Noisy Student Training pipeline for ASR}
\label{fig:nst-pipeline}
\end{figure}

Noisy Student Training \cite{NST_ASR2_pushing_limits} is an iterative self-training method evolved from Teacher Student Learning, the pipeline of which is illustrated in Fig \ref{fig:nst-pipeline}. 
Initially a teacher model is trained with supervised data and pseudo-labels are generated by the teacher. Then data augmentation methods such as SpecAug \cite{park19e_interspeech} and speed perturbation are applied during training and the student model is trained using both augmented supervised data and pseudo-data. 

In our pipeline, we follow the design that the student model always has same parameters as the teacher, and adopts dropouts and stochastic depths so that the student could be more robust and general than the teacher when it is trained.
After training finishes, the student model will be assigned as the new teacher and the whole pipeline will iterate. 
After several rounds of training, models trained to tolerant noises and augmentations will tend to have better performance generally.

\section{Data Selection Strategy}
\begin{CJK*}{UTF8}{gbsn}

Data selection and filtering play a significant role in SSL especially in an out-of-domain situation. This circumstance occurs frequently in the industry when we have limited labelled data in the desired area or in low resource tasks. 

Initially, standard NST is performed on AISHELL-1 as supervised data and AISHELL-2 as unsupervised data without any other filtering strategy. The generated pseudo labels have quite promising results of 8.38\% CER but when unsupervised data is set to WenetSpeech which has different domain and recording settings, pseudo-label’s CER increases dramatically to 47.1\% which is unacceptable for training. 
\textit{LM Filter} is then proposed to improve the performances of NST when non-target domain data is provided.

\begin{figure}[htb]
\centering
\includegraphics[width=8.5 cm]{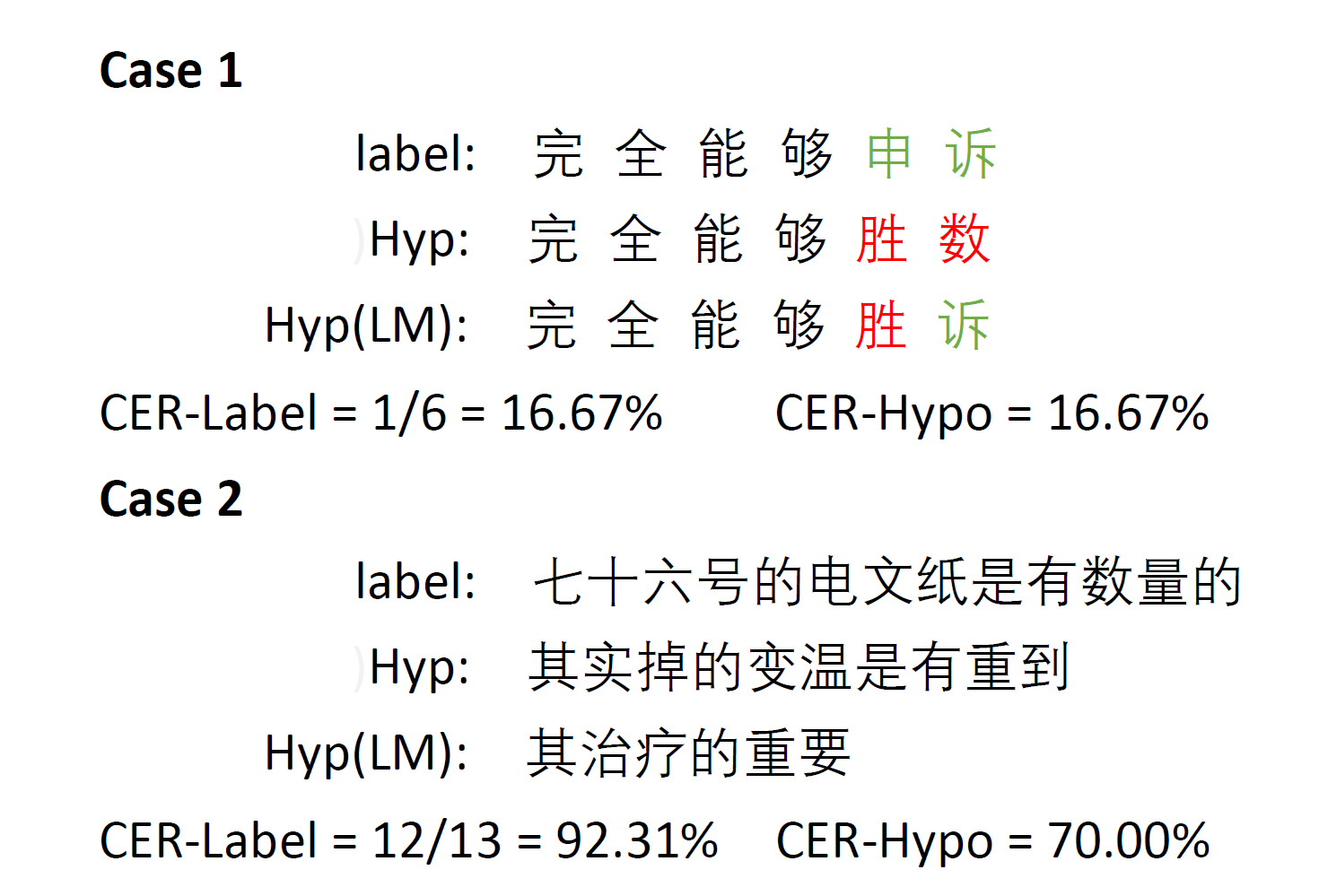}
\caption{Examples of how to calculate the CER-Label and CER-hyp of sentences.}
\label{fig:cer-hyp}
\end{figure}

Our hypothesis is that if a language model believes the sentence does not require any further modification, then this sentence has higher probability of being a correct pseudo-label. Here we introduce two definitions and examples to better understand how our \textit{LM Filter}  works. 
\begin{itemize}
    \item \textbf{CER-Hypo} is the CER between student model’s hypothesis with greedy decoding and student model’s hypothesis with Language model.
    \item \textbf{CER-Label} is the CER between student model’s hypothesis with Language model and the true label.
\end{itemize}
We evaluated our method on the Mandarin corpus using CER while the same definition can be applied to other languages e.g. English by replacing CER with WER. Two cases are listed in Fig \ref{fig:cer-hyp}.
In case 1, the difference between the hypothesis with greedy decoding and the hypothesis decoding with LM is 1 character (eg. char “数” and char “诉”) so the CER-Hypo is 16.67\% . The CER-Label is also 16.67\% in this case, since it takes 1 substitution step to transfer the hypothesis to true label (eg. char “申” and char “胜”). 
In case 2, the sentence is more challenging than the first case for the initial student model. 
The student model learns partial acoustic features but the transcripts are mostly wrong. 
The LM tends to make more modifications due to the low probabilities of such sentence in the corpus. The CER-Hypo and CER-Label both are extremely high in this case.


A large amount of cases suggest that \textbf{CER-Hypo} and \textbf{CER-Label} have strong positive correlations, sentences with lower CER-Hypo tend to have lower CER-Label. Our \textit{LM Filter} uses CER-Hypo as a threshold (eg.10\%) to filter out high CER-Label data. We also observe that unsupervised data with similar domain to supervised data are more likely to have lower CER-Hypo values.  For unsupervised data from Youtube, similar topics in ``readings"and ``news" tend to have lower CER-Hypo and non-target domain such as ``drama" and ``variety" are more likely to be removed by the \textit{LM Filter}. We also propose a speaking rate filter for WenetSpeech dataset, which is the hypothesis length divided by audio time. The music and song audios that are common elements of drama and variety shows can be effectively removed by this filter.

\end{CJK*}



\section{Experiments and results}
\label{Experiment}

\subsection{Datasets and domain description}
\label{ssec:datasets}

We evaluate our proposed data selection strategy on the following three datasets: AISHELL-1, AISHELL-2 and WenetSpeech.
AISHELL-1 is a 178-hour open-sourced Mandarin speech corpus, with strictly annotated and inspected transcriptions which mainly covers 5 topics of Finance, Technology, Sports, Entertainments and News.
AISHELL-2 consists of 1k hours of Mandarin speech with the same device and recording environment settings as AISHELL-1. The major topics of these two datasets are similar, but the transcripts and audios of the test set are different.
WenetSpeech has 10k hours of speech where transcripts are generated by OCR on video data from Youtube and Podcast, which lacks inspection and accuracy. Domains are diverse and mostly consists of Drama, Variety show and Audio books.

\subsection{Experiment settings}
\label{ssec:settings}
First, we use AISHELL-1 as the supervised dataset and treat AISHELL-2 and WenetSpeech as unsupervised data. Initially 1k hours of WenetSpeech data are randomly selected to match the size of AISHELL-2. 
And then the size of WenetSpeech data is increased up to 4k hours to test the degree of saturation for unsupervised data. Eventually, we switch the supervised dataset to AISHELL-2 to evaluate the performances of our data selection strategy on industrial-level supervised datasets. The upper bound of data ratio for supervised and unsupervised data is set to 1:9.

The neural structures for both teacher and student models are the same, which is a 16-layer Conformer model \cite{conformer}. 
Our language model is a 5-gram model with corpus contains training texts as well as extra wiki texts.
All experiments are conducted in WeNet toolkit \cite{yao2021wenet} and NVIDIA A100 GPUs.
We perform 7 iterations of NST with and without data selection strategy on WenetSpeech and 5 iterations on AISHELL2.

%

\subsection{Baselines}
Supervised baseline using only AISHELL-1 data, which is the initial teacher of NST iterations is shown in Table \ref{table:s-baselines}. Then supervised training is done on AISHELL-1 data mixing with supervised AISHELL-2 and WenetSpeech. These two results are considered as ceilings of our model’s performance. Then standard NST experiments is conducted without data selection strategy using AISHELL-1 as supervised data and AISHELL-2 as unsupervised data, the results of which are shown in Table \ref{table:s-baselines}. 
The 3.99\% CER can be achieved after first NST iteration because these two datasets have similar domain and recording settings. The closer the topics and configs, the better performance the NST algorithm will have. In the case of the ideal data distribution, the filtering approach is not required. However, in the majority of recognition jobs, this condition is not typical.
After first NST iteration with WenetSpeech pseudo-label, the CER increases to 5.52\%, which is even higher than the supervised baseline using only AISHELL-1 data. To reduce CER of pseudo-labels and make training easier in early stages, an appropriate filter is required.

\begin{table}[]
\centering
\caption{CER for supervised baselines and standard NST first iteration with AISHELL-2 and 1k WenetSpeech dataset. }
\label{table:s-baselines}
\begin{tabular}{SSS} \toprule
    {Supervised} &{Unsupervised}& {Test CER}    \\ \midrule
    {AISHELL-1 Only} & {---}  & 4.85  \\
    {AISHELL-1 + WenetSpeech}& {---} & 3.54   \\
    {AISHELL-1 + AISHELL-2}  & {---} & 1.01 \\
    {AISHELL-1} & {WenetSpeech}  & 5.52  \\
    {AISHELL-1}& {AISHELL-2} & 3.99   \\

\bottomrule
\end{tabular}
\end{table}


\subsection{Data selection strategy performances}
\label{ssec:Experiment Results}


Performances of \textit{LM Filter} of supervised AISHELL-1 data and unsupervised WenetSpeech data are shown in Table \ref{table:nst-aishell-1}. 
The best 4.31\% CER can be achieved after 7 iterations in the test set. There can be relatively 11.13\% CER reduction compared with the supervised baseline. In addition to the test set's CER, the following three metrics are used to assess the quality of the pseudo-label : \textit{Pseudo CER} which is referred to the CER of pseudo-labels, \textit{Filtered CER} and \textit{Filtered hours} which are the CER and the duration of filtered unsupervised data.

Results in initial iteration indicate that the \textit{LM Filter} can significantly decrease the Pseudo CER from 47.1\% to 25.18\% which makes the pseudo-label satisfactory for further training.  
The Pseudo CER and Filtered CER decrease as the number of iterations rises, and \textit{LM Filter} permits more filtered data to be fed into the model. This suggests that \textit{LM Filter} may gradually learn noisy information, and our student model could make even greater use of non-target domain data.

\begin{table}
\caption{Performances of \textit{LM Filter} on supervised AISHELL-1 data and unsupervised 1k WenetSpeech data, including the CER of test set, Pseudo CER, Filtered CER and Filtered hours.   }
\label{table:nst-aishell-1}
\begin{tabular}{p{10mm}<{\centering} p{20mm}<{\centering} p{10mm}<{\centering} p{10mm}<{\centering} p{10mm}<{\centering}} \toprule
    {\# NST Iter} & {AISHELL-1 test CER} & {Pseudo CER} & {Filtered CER} & {Filtered hours}     \\ \midrule
    0   & 4.85 & 47.10 & 25.18 & 323 \\
    1   & 4.86 & 37.02 & 20.93 & 436 \\
    2   & 4.75 & 31.81 & 19.74 & 540 \\
    3   & 4.69 & 28.27 & 17.85 & 592 \\
    4   & 4.48 & 26.64 & \textbf{14.76} & 588 \\
    5   & 4.41 & 24.70 & 15.86 & 670 \\
    6   & 4.34 & \textbf{23.64} & 15.40 & 669 \\
    7   & \textbf{4.31} & 23.79 & 15.75 & \textbf{694} \\ \bottomrule
\end{tabular}
\end{table} 


\begin{figure*}
    \centering
    \includegraphics[width=1.0\textwidth]{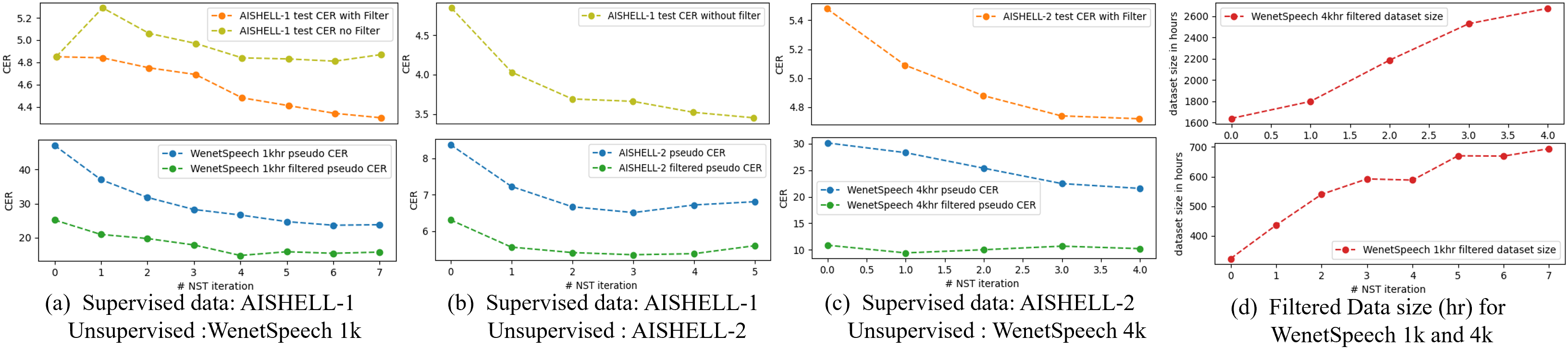}
    \caption{This figure illustrates results of NST with our data selection strategy using different supervised and unsupervised data. CER performances in test sets of standard NST without \textit{LM filter} are marked in yellow lines and with \textit{LM filter} are in orange lines. Pseudo CERs are shown in blue lines and green lines gives the Filtered CER. Filtered hours of unsupervised data are shown in (d). }
    \label{fig:nst-all}
\end{figure*}

\subsection{Discussions}

\textbf{Multiple NST iterations}: Multiple NST iterations are conducted to show our proposed data selection strategy can achieve better performance and converge faster. Fig \ref{fig:nst-all} displays all the results of AISHELL-1.
When WenetSpeech is used as unsupervised data in section (a), after 7 iterations of NST without filter strategy, a negligible improvement is obtained. In contrast, \textit{LM Filter} can yield a relative improvement of 10.4\% with faster training time.
With our \textit{LM Filter}, CERs for test set and pseudo-label drop gradually and the filtered data size grows during each training iterations.
In section (b), when using AISHELL-2 as unsupervised data, 3.45 \% CER can be achieved after the 5 NST training without filter. The relatively small initial Pseudo CER of 8.38 \% in AISHELL-2 indicates that unsupervised data with matched domain can generate effective pseudo-labels to acquire the requirement of NST training. Additionally, we perform extra iteration that combines all pseudo-labels that have been filtered by final NST models on both WenetSpeech and AISHELL-2, yielding the best CER result of 3.31\%. According to our understanding, this is the best current result in AISHELL-1 test set without any further supervised data.


\textbf{Impact of domains}: Our experiments indicate that NST approach is very sensitive to the domain issue. 
Domain can have a significant impact on the effectiveness of the NST algorithm. 
The quality of pseudo-labels tends to rise if WenetSpeech samples are taken from tags that are more closely related to the AISHELL domain (such as Readings and News). In contrast, tags for drama and variety show that are not commonly used in AISHELL yielded inferior pseudo-labels.
The pseudo-labels' quality will further affect the filtered data size and NST iterations' converging speed. 
Among all the topics, we also discover that sources like Audio books and Podcast most likely provide pseudo-labels with higher qualities.

\textbf{Effectiveness on large dataset}: To further demonstrate our \textit{LM Filter}’s effectiveness on large supervised dataset, we conduct experiments using AISHELL-2 as supervised data. The CER results are shown in Table \ref{table:nst-aishell-2}.
In the AISHELL-2 test set, 13.6\% relative improvement is achieved, which further demonstrates \textit{LM Filter}’s scalability on larger supervised data under industrial scale. Detail performances is shown in plot (c) of Fig \ref{fig:nst-all}, it illustrates similar trends as AISHELL-1. 

\begin{table}
\caption{Results of AISHELL-2 test set when using supervised AISHELL-2 data and unsupervised 4k hr WenetSpeech data after applying \textit{LM Filter} . }
\label{table:nst-aishell-2}
\begin{tabular}{p{10mm}<{\centering} p{20mm}<{\centering} p{15mm}<{\centering} p{10mm}<{\centering} p{10mm}<{\centering}} \toprule
    { \# NST Iter} & { AISHELL-2 test CER} & {Pseudo CER} & {Filtered CER} & {Filtered hours}    \\ \midrule
    0   & 5.48 & 30.10 & 11.73 & 1637 \\
    1   & 5.09 & 28.31 & \textbf{9.39} & 2016 \\
    2   & 4.88 & 25.38 & 9.99 & 2186 \\
    3   & 4.74 & 22.47 & 10.66 & 2528 \\
    4   & \textbf{4.73} & \textbf{22.23} & 10.43 & \textbf{2734} \\
\bottomrule
\end{tabular}

\end{table}

\section{Conclusions}
\label{sec:majhead}

In this paper, a novel data selection strategy named \textit{LM Filter} is proposed to improve the performances of NST in non-target domain data, which utilizes the model differences from decoding strategies. 
Results reveal significant improvements of 10.4\% compared to baselines with no data filtering. we obtain 3.31\% CER in AIShell-1 test set, which is best result according to our knowledge without any further supervised data. In addition, we perform evaluations on 1k hour AIShell-2 dataset and achieve 4.73\% CER on test set, which further demonstrates the robustness of  \textit{LM Filter} with larger supervised data.




\ninept
\bibliographystyle{IEEEbib}
\bibliography{strings}

\end{document}